Research article

JHEP|Reports# Primary liver cancer classification from routine tumour biopsy using weakly supervised deep learning

Authors
**Aurélie Beaufrère, Nora Ouzir,** Paul Emile Zafar, Astrid Laurent-Bellue, Miguel Albuquerque, Gwladys Lubuela, Jules Grégory, Catherine Guettier, Kévin Mondet, Jean-Christophe Pesquet, Valérie Paradis

Correspondence
aurelie.beaufrere@aphp.fr (A. Beaufrère).Graphical abstract

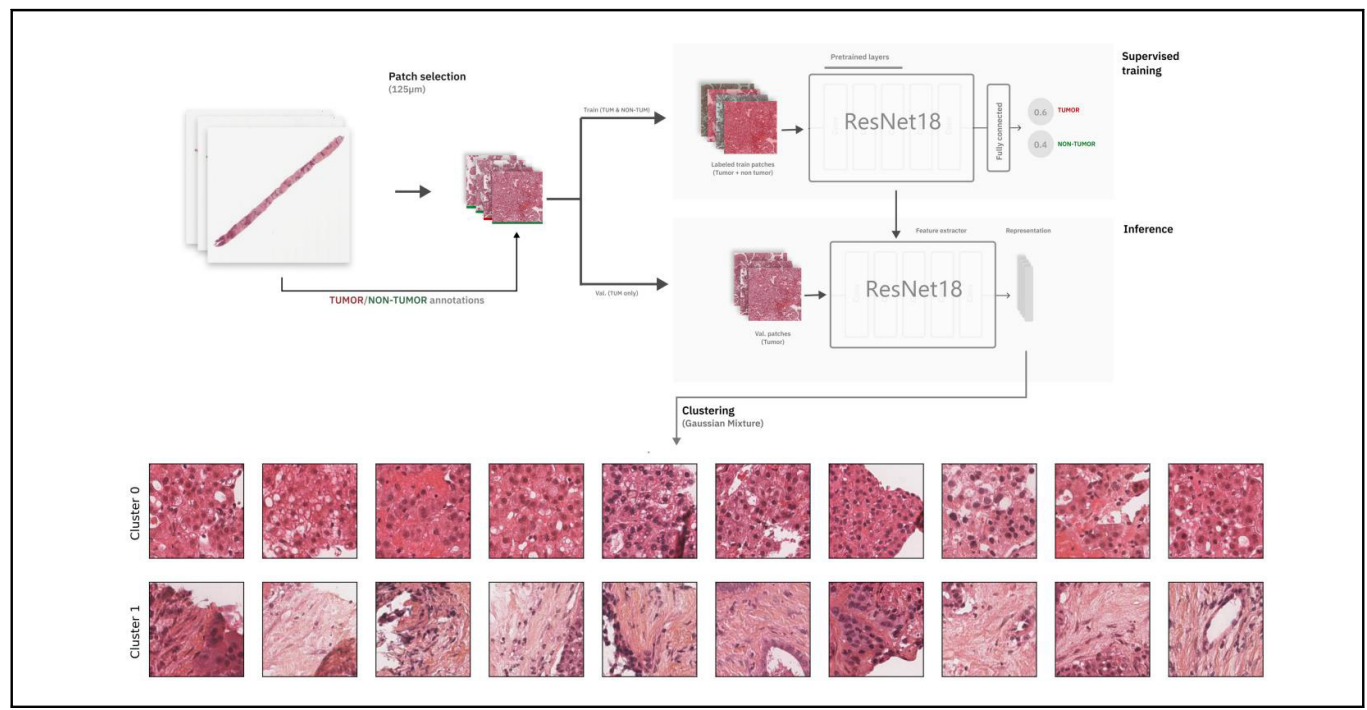

## Highlights

- Weakly supervised learning method identified specific features of HCC and iCCA.
- No specific features of cHCC-CCA were recognized.
- Our model allowed for assessment of the proportion of HCC and iCCA tiles within a cHCC-CCA biopsy.
- Our model correctly predicted the diagnosis in 97% of HCC cases and 83% of iCCA cases.
- For cHCC-CCA, we observed a highly variable proportion of HCC and iCCA tiles.

## Impact and implications

The diagnosis of primary liver cancers can be challenging, especially on biopsies and for combined hepatocellular-cholangiocarcinoma (cHCC-CCA). We automatically classified primary liver cancers on routine-stained biopsies using a weakly supervised learning method. Our model identified specific features of hepatocellular carcinoma and intrahepatic cholangiocarcinoma. Despite no specific features of cHCC-CCA being recognized, the identification of hepatocellular carcinoma and intrahepatic cholangiocarcinoma tiles within a slide could facilitate the diagnosis of primary liver cancers, and particularly cHCC-CCA.

https://doi.org/10.1016/j.jhepr.2024.101008

# Primary liver cancer classification from routine tumour biopsy using weakly supervised deep learning

Aurélie Beaufrère,[1,2,3,*,†] Nora Ouzir,[4,†] Paul Emile Zafar,[1,4] Astrid Laurent-Bellue,[5] Miguel Albuquerque,[1] Gwladys Lubuela,[2] Jules Grégory,[1,2,6] Catherine Guettier,[5] Kévin Mondet,[1] Jean-Christophe Pesquet,[4] Valérie Paradis[1,2,3]

[1]AP-HP. Nord, Department of Pathology, FHU MOSAIC, Beaujon Hospital, Clichy, France; [2]Université Paris Cité, Paris, France; [3]Centre de Recherche sur l'Inflammation, INSERM UMR 1149, Paris, France; [4]University of Paris-Saclay, CentraleSupélec, CVN, OPIS Inria, Gif-sur-Yvette 91190, France; [5]AP-HP, Department of Pathology, Hôpital Bicêtre, Le Kremlin- Bicêtre, France; UMR-S 1193, Université Paris-Saclay, Kremlin-Bicêtre, France; [6]AP-HP.Nord, Department of Imaging, Beaujon Hospital, Clichy, France

**Background & Aims:** The diagnosis of primary liver cancers (PLCs) can be challenging, especially on biopsies and for combined hepatocellular-cholangiocarcinoma (cHCC-CCA). We automatically classified PLCs on routine-stained biopsies using a weakly supervised learning method.
**Method:** We selected 166 PLC biopsies divided into training, internal and external validation sets: 90, 29 and 47 samples, respectively. Two liver pathologists reviewed each whole-slide hematein eosin saffron (HES)-stained image (WSI). After annotating the tumour/non-tumour areas, tiles of 256x256 pixels were extracted from the WSIs and used to train a ResNet18 neural network. The tumour/non-tumour annotations served as labels during training, and the network's last convolutional layer was used to extract new tumour tile features. Without knowledge of the precise labels of the malignancies, we then applied an unsupervised clustering algorithm.
**Results:** Pathological review classified the training and validation sets into hepatocellular carcinoma (HCC, 33/90, 11/29 and 26/47), intrahepatic cholangiocarcinoma (iCCA, 28/90, 9/29 and 15/47), and cHCC-CCA (29/90, 9/29 and 6/47). In the two-cluster model, Clusters 0 and 1 contained mainly HCC and iCCA histological features. The diagnostic agreement between the pathological diagnosis and the two-cluster model predictions (major contingent) in the internal and external validation sets was 100% (11/11) and 96% (25/26) for HCC and 78% (7/9) and 87% (13/15) for iCCA, respectively. For cHCC-CCA, we observed a highly variable proportion of tiles from each cluster (cluster 0: 5-97%; cluster 1: 2-94%).
**Conclusion:** Our method applied to PLC HES biopsy could identify specific morphological features of HCC and iCCA. Although no specific features of cHCC-CCA were recognized, assessing the proportion of HCC and iCCA tiles within a slide could facilitate the identification of cHCC-CCA.
**Impact and implications:** The diagnosis of primary liver cancers can be challenging, especially on biopsies and for combined hepatocellular-cholangiocarcinoma (cHCC-CCA). We automatically classified primary liver cancers on routine-stained biopsies using a weakly supervised learning method. Our model identified specific features of hepatocellular carcinoma and intrahepatic cholangiocarcinoma. Despite no specific features of cHCC-CCA being recognized, the identification of hepatocellular carcinoma and intrahepatic cholangiocarcinoma tiles within a slide could facilitate the diagnosis of primary liver cancers, and particularly cHCC-CCA.

## Introduction

Primary liver cancers (PLCs) are the third leading cause of cancer-related death worldwide, with an increasing incidence in Western countries.[1,2] PLCs define a heterogeneous group of tumours associated with distinct risk factors, clinical findings, imaging, and histologic and molecular characteristics. Among these tumours, hepatocellular carcinoma (HCC) and intrahepatic cholangiocarcinoma (iCCA) are the most common and represent the two ends of the PLC spectrum. Halfway between HCC and iCCA, combined hepatocellular-cholangiocarcinoma (cHCC-CCA), a much rarer tumour, shares features with both types.[3–5]

The diagnosis of HCC is mainly based on imaging in the context of cirrhosis, whereas the gold standard for diagnosing iCCA and cHCC-CCA is histological analysis.[6–8] The pathological definition of cHCC-CCA has significantly evolved. The last 2019 WHO classification endorsed a cHCC-CCA diagnosis definition based on the unequivocal presence of both hepatocytic and cholangiocytic differentiation within the same tumour on routine histopathology (hematein eosin or hematein eosin saffron [HES] staining). Additional immunostainings can help confirm the diagnosis.[9] Nonetheless, diagnosing PLCs, particularly cHCC-CCA, is challenging on biopsy samples because histological analysis may miss one tiny tumour component. Our previous study concluded that although diagnostic performance remains relatively low for cHCC-CCA on routine biopsy staining, additional immunohistochemistry (IHC) and, mainly, a two-step strategy combining imaging and histology, significantly improved the diagnostic performance of biopsy for cHCC-CCA.[10]

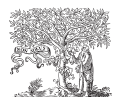

Histology slides are readily available in pathology departments, and the recent advance of high-throughput scanning devices has enabled digital captures of an entire conventional glass slide, called a whole-slide image (WSI).[11,12] WSIs contain valuable information that can be invisible to the human eye. Artificial intelligence (AI), particularly deep-learning methods, can exploit this information at different levels by extracting deep morphological features from WSIs after a tiling step.[13,14] The literature on AI applications has flourished in recent years. Most studies in computational pathology rely on supervised methods, where ground truth is known only for the slide (globally) and hardly ever for every single tile.[15] The effort of an exhaustive annotation has been a significant obstacle to many supervised AI methods in medical applications. In histology, in which the image size is huge, no reasonable human effort will likely succeed in labelling a massive dataset (at the tile scale) for fully supervised AI methodologies.[16]

Weakly supervised methods circumvent this difficulty by using a feasible amount of annotations, and several studies have already shown the effectiveness of these methods in extracting meaningful features for medical applications, including histology.[16,17] The essence of the solved task remains unsupervised, although not entirely blind to the application specificities anymore. In imaging applications, this method could help provide a few categorical labels for WSI segmentation[16] or a limited number of classes to solve a more complex and broad classification problem.[18] In liver pathology, weak annotations of HCC histological slides have been used to extract features for prognosis prediction.[19–21] Different AI models trained on histology slides of HCC treated by surgical resection showed that weakly supervised models outperformed others in predicting the activation of immune and inflammatory gene signatures.[14] Although some of these studies have investigated weakly supervised methods on WSIs from surgical specimens, biopsy studies remain scarce.

This proof-of-concept study aimed to automatically classify PLCs from biopsy samples using a weakly supervised deep-learning approach.

## Materials and methods
### Pathological analysis
*Sample selection*
We selected 119 formalin-fixed paraffin-embedded biopsies of PLC archived between 2012 and 2021 in the Pathology Department of Beaujon Hospital (Clichy, France), divided into 90 training and 29 internal validation samples. Both sets comprised a balanced proportion of HCC, iCCA and cHCC-CCA to train and validate the model correctly. We constructed an external validation set of 47 biopsies of PLC archived in the Pathology Department of Bicêtre Hospital between 2012 and 2023 (Le Kremlin-Bicêtre, France) (Fig. 1). All patients gave written consent for using the biopsies as required by French legislation. This study was registered at the Commission Nationale de l'Informatique et des Libertés and was approved by the ethics committee (EDS APHP no. CSE-20-85 MOSAIC-EDS). The recorded data included age, sex, presence of surgery, main risk factors for chronic liver diseases (*e.g.*, viral hepatitis, chronic alcohol consumption, metabolic syndrome, hemochromatosis), Child-Pugh score, METAVIR fibrosis score in the non-tumour liver, and the number and size of tumours at diagnosis.

*HES and IHC*
For each case, an HES staining was selected for digitization. An additional IHC analysis was performed for cases included in the internal validation set when sufficient material was available (20/29 cases). Whole-biopsy sections were cut on charged slides (Superfrost slides; Thermo Scientific, Waltham, MA, USA) and were immunostained for routine markers for HCC and iCCA (glypican 3 [1G12, 1/100, Zytomed, Germany], anti-hepatocyte antibody [OCH1E5, 1/250, Agilent, USA], and CK7 [50V-TL 12/30, 1/500, Agilent, USA]). HES and IHC slides were scanned at x20 magnification (Aperio Slide Scanner).

*Pathology review*
Two liver pathology experts (AB and VP) examined all HES WSIs, classifying each case as HCC, iCCA, or cHCC-CCA (Fig. 2). cHCC-CCA diagnosis was retained when biphenotypic differentiation (hepatocytic and cholangiocytic) was evident on HES WSIs according to the definition of the last WHO classification.[9]

*Quantitative IHC analysis*
Quantitative IHC analysis was performed in the internal validation set to improve the assessment of the proportion of iCCA and HCC contingents, particularly for cHCC-CCA. Single-cell-based analyses were carried out on validation WSIs using the haematoxylin channel to segment cell nuclei in the open-source WSI analysis software QuPath (v0.2.3).[22] A stepwise procedure followed segmentation for further cell subclassification. First, cells were classified as tumourous or stromal cells using QuPath's built-in machine-learning features as described.[23] Tumour and stromal cells were further subclassified using intensity thresholds in the relevant chromogen channels (DAB). This subclassification comprised the identification of glypican 3-, anti-hepatocyte antibody, and CK7-positive cells within tumour areas (Fig. 3). A ratio of positive tumour cells to tumour cells was calculated for each immunostaining. The percentage of HCC IHC contingent was determined with the glypican 3 or anti-hepatocyte antibody ratio (the highest positive ratio between the two antibodies was retained). The percentage of iCCA IHC contingent was determined with the CK7 ratio. We used the Pearson correlation coefficient to assess the diagnostic agreement between the AI model and pathology.

### Weakly supervised tumour classification
*Patch processing*
Processing WSIs is computationally challenging because of their high dimension (∼$10^4$ pixels wide, which results in images of 100 to 1,000 megapixels). Because a straightforward resizing of the WSIs can lead to loss of crucial information at the microscopic scale, we adopted a patch-wise strategy, with patches (or tiles) of a fixed, much smaller size extracted from the images and processed individually. After testing different scales (*i.e.*, patch sizes), we chose a scale of 125 μm, corresponding to a patch size of 256x256. This choice allowed us to study the samples at an adequate level of tissue detail while guaranteeing a reasonable computational time. Because tissue areas are scarce within biopsies, this procedure also maximized the number of extracted patches compared to choosing a larger patch size. Large blank background areas cover the rest of the images; they were processed using a masking approach, ignoring the background and extracting patches lying on the tissue areas only (along the shape of the biopsy). We further processed border patches (*i.e.,* patches containing background and tissue) by selecting the ones containing mainly tissue pixels.

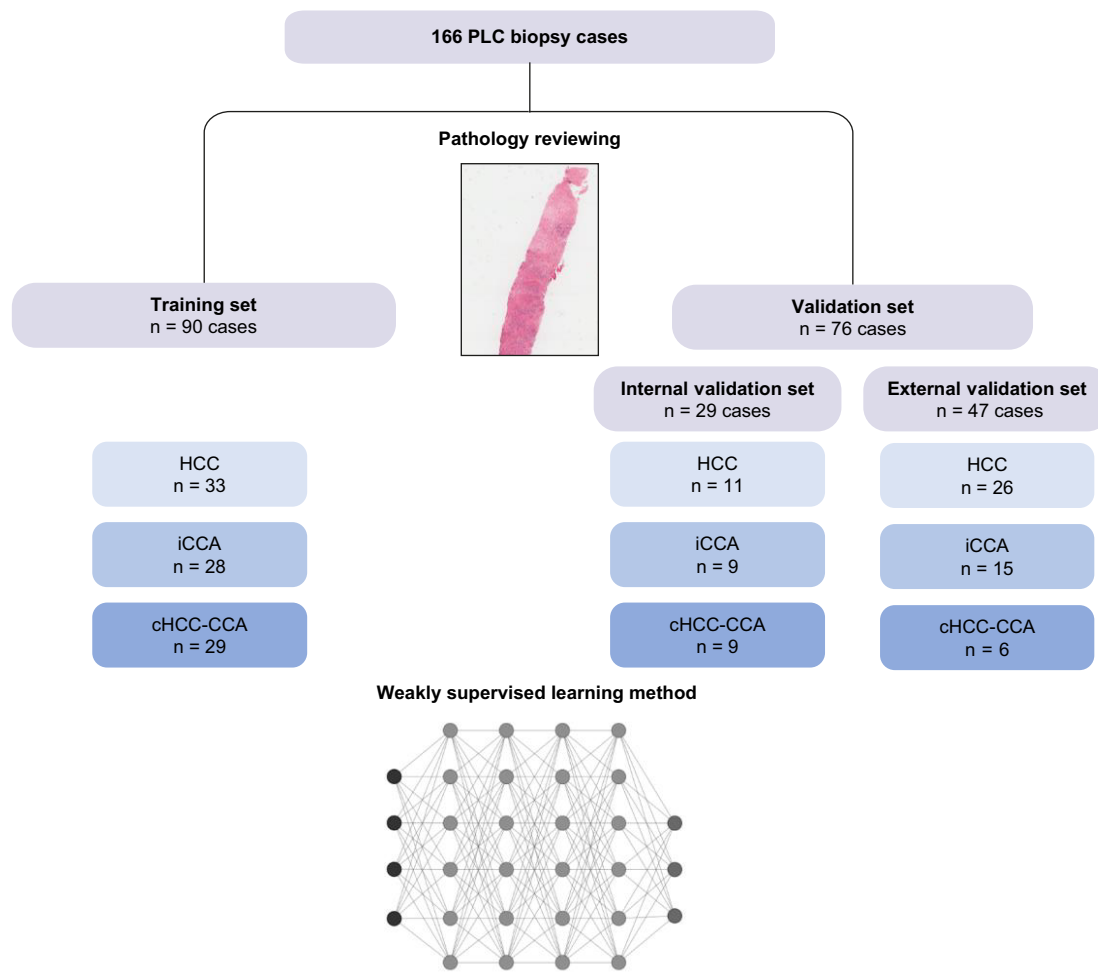

**Fig. 1. Flowchart of the study.** We selected 166 formalin-fixed paraffin-embedded biopsies of PLC, divided into 90 training, 29 internal validation and 47 external validation samples. cHCC-CCA, combined hepatocellular-cholangiocarcinoma; HCC, hepatocellular carcinoma; iCCA, intrahepatic cholangiocarcinoma; PLC, primary liver cancer.

*Data augmentation*
After patch extraction, we further augmented the data to bring more variability to the dataset. The data augmentation was restricted to plausible transformations to preserve meaningful patches from the pathological point of view: colour variations (±2% hue, [-80% to +60%] saturation, and ±20% brightness), rigid transformations such as horizontal and vertical flips, random rotations of (-90° to +90°), and Gaussian blurring with a kernel of size 3x3. We varied the 125 μm scaling by a small factor between 1 and 1.2 to add variability while preserving similar structures in the patches. The final number of patches was 86,936, including 63,132 (∼70%) training patches and 23,804 (∼30%) validation patches. To avoid over-fitting, training and validation patches were extracted from biopsies of distinct patients. Both sets contained a balanced proportion (∼1/3) of patients with each type of cancer (Fig. 1).

*Tumour/non-tumour annotations*
The weakly supervised approach involved using tumour/non-tumour annotations, which are more easily accessible than detailed tumour-type annotations. The experts manually outlined the tumoural regions, and their annotations were transformed into a binary mask of tumour/non-tumour regions inside/outside the outlined area. Border patches were considered tumoural if most pixels were labelled as such. All biopsies of training and validation cohorts contained tumoural and non-tumoural tiles. We used both tumoural and non-tumoural tiles for training but only tumoural tiles for validation.

*Feature extraction and clustering*
The proposed deep learning-based approach involved a three-step process (Fig. 4). First, supervised learning trained a feature extractor guided by the tumour/non-tumour annotations. The feature extractor was applied to tumoural validation patches in the second phase, transforming them according to the learned representation. Finally, an unsupervised learning method clustered the transformed validation patches into $K = 2$ or $K = 3$ groups. We used principal component analysis to reduce the number of features by half before clustering.

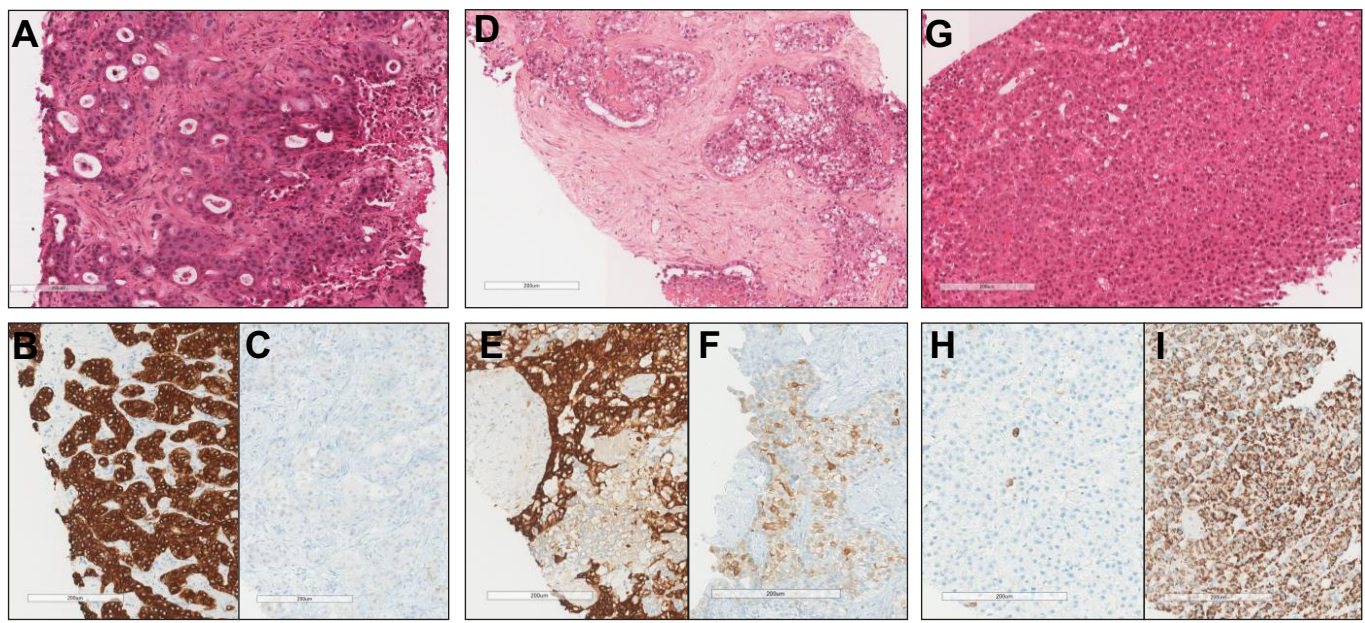

**Fig. 2. Examples of morphological and immunohistochemistry features of iCCA, cHCC-CCA and HCC.** (A) iCCA, HES staining: glandular architecture with fibrous stroma. (B) iCCA, CK7 staining: intense labelling of tumour cells. (C) iCCA, glypican 3 staining: no labelling of tumour cells. (D) cHCC-CCA, HES staining: glands surrounded by cuboidal tumour cells mixed with trabeculae and nests of clear hepatoid tumour cells. (E) cHCC-CCA, CK7 staining: expression of the majority of tumour cells. (F) cHCC-CCA, glypican 3 staining: expression of tumour cells located in the trabeculae and the nests. (G) HCC, HES staining: hepatoid tumour cells organized in trabeculae and pseudoglands without fibrous stroma. (H) HCC, CK7 staining: rare tumour cells marked. (I) HCC, anti-hepatocyte antibody staining: intense expression of all of the tumour cells. cHCC-CCA, combined hepatocellular-cholangiocarcinoma; HCC, hepatocellular carcinoma; HES, hematein eosin saffron; iCCA, intrahepatic cholangiocarcinoma.

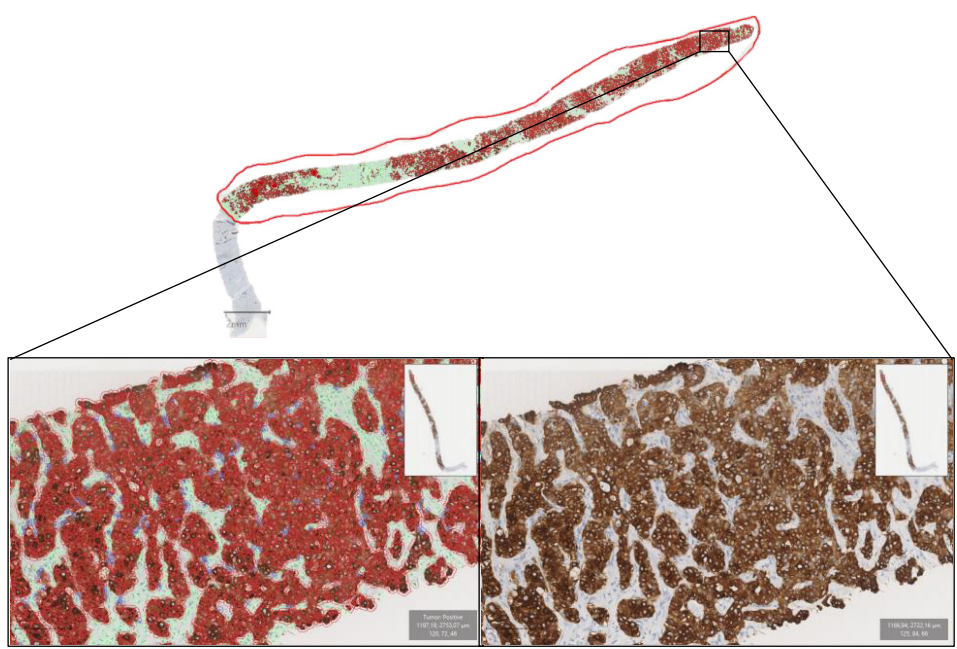

**Fig. 3. Quantitative immunohistochemistry analysis with QuPath software.** Example analysis of CK7 staining in an iCCA case. The tumour area was manually annotated. Cells were classified as tumour cells (red or blue) or stromal cells (green) by using the QuPath machine-learning features. Then CK7-positive tumour cells (red) were further subclassified by using intensity thresholds in the tumour area relevant chromogen channels (DAB). iCCA, intrahepatic cholangiocarcinoma.

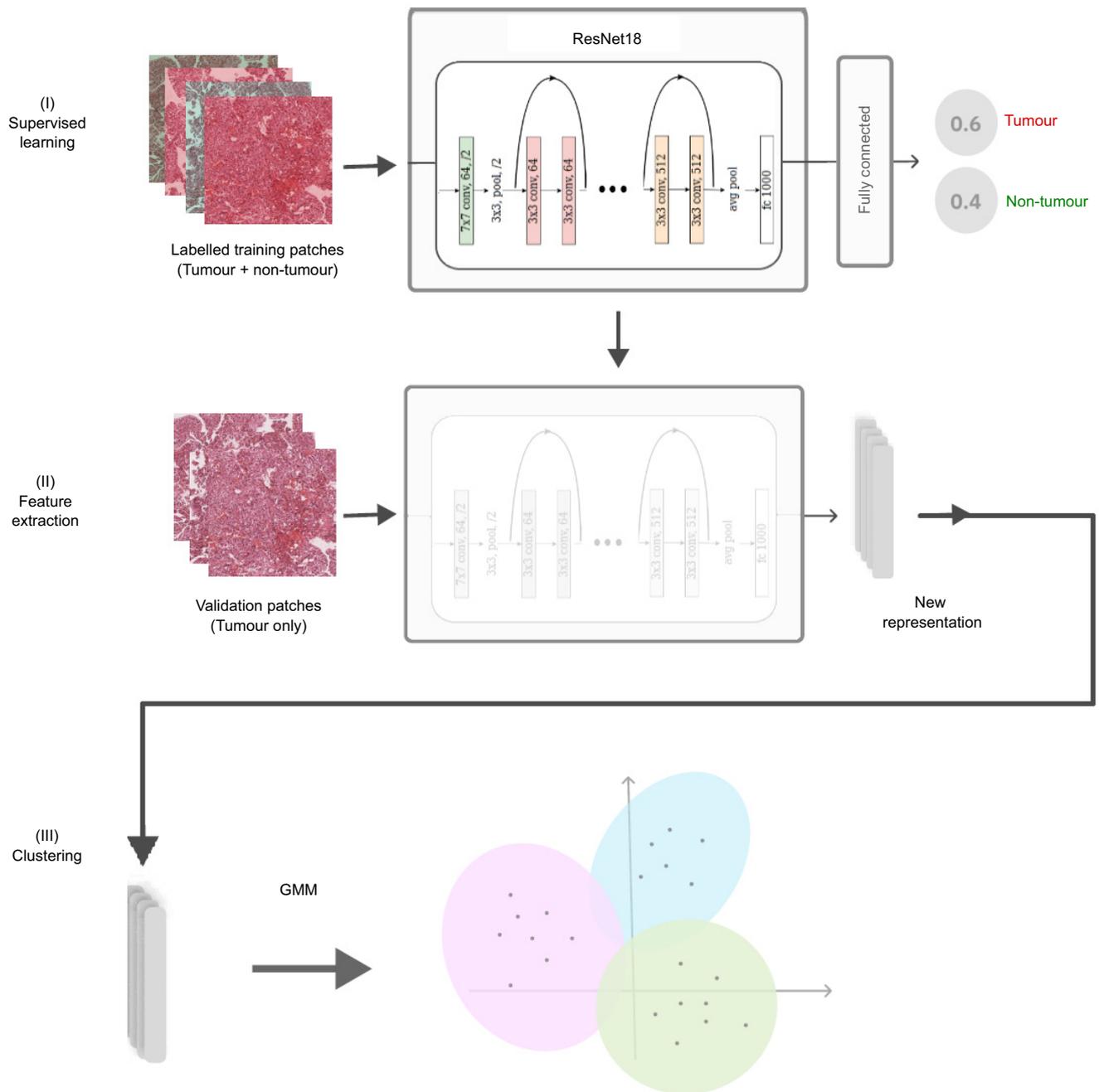

**Fig. 4. Main steps of the proposed weakly supervised method.** After annotating the tumour/non-tumour areas, tiles of 256x256 pixels were extracted from the whole-slide images and used to train a ResNet18 neural network (step I). The tumour/non-tumour annotations served as labels during training, and the network's last convolutional layer was used to extract new tumour tile features (step II). Without knowledge of the precise labels of the malignancies, we then applied an unsupervised clustering algorithm (GMM) (step III). GMM, Gaussian mixture model.

*Implementation details*
The first stage of the method used a pre-trained ResNet18 architecture that was fine-tuned on its last eight convolutional layers with the annotated 63,132 training patches. The weights of the first nine layers pre-trained on ImageNet[24] were unchanged. The ResNet18 architecture, although less complex than its deeper 50- and 150-layer versions,[25] provided a solid baseline for assessing the potential of transfer learning for our task. Recent works using models pre-trained on ImageNet have demonstrated that more complex architectures did not necessarily lead to clear improvements and significantly increased the computational load.[26] Hence, ResNet18, with fewer parameters, is less prone to over-fitting and over-specialization.

We used an Adam optimizer with an initial learning rate of 0.03 and an exponential decay scheduler with a multiplicative factor of 0.5. The obtained training and validation accuracies were 0.75 and 0.71, respectively. After training, the final dense level of the network was discarded. The last convolutional layer was used to extract the features from the 23,804 validation patches, which led to a new patch representation of size 256x256x512 (*i.e.*, 512 features representing each pixel). After principal component analysis, the patch size was reduced to 256x256x256. The unsupervised clustering algorithm was the Gaussian mixture model, which was solved with the expectation-maximization algorithm.[27] After clustering, the WSI can be reconstructed to visualize the spatial distribution of the tumour labels (Fig. 5).

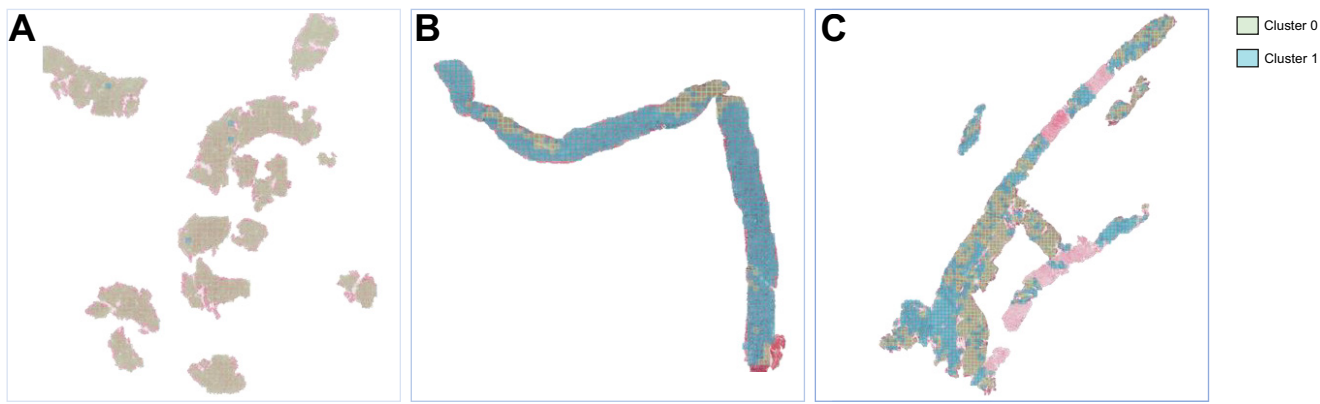

**Fig. 5. Spatial distribution of the tumour labels for the two-cluster model in examples of PLCs.** Examples of HCC (A), iCCA (B), and cHCC-CCA patients (C). cHCC-CCA, combined hepatocellular-cholangiocarcinoma; HCC, hepatocellular carcinoma; iCCA, intrahepatic cholangiocarcinoma.

## Results

### Clinical and pathological features of cases

Clinical features of cases included in the two datasets are summarized in Table S1. Pathological review classified the training, internal and external validation sets into HCC (33/90, 11/29 and 26/47, respectively), iCCA (28/90, 9/29 and 15/47, respectively), and cHCC-CCA (29/90, 9/29 and 6/47, respectively) (Fig. 1). In the internal validation set, all cHCC-CCA (n = 9) cases presented a majority contingent of iCCA. The mean percentage of iCCA IHC contingent was 90% (range 79-99) in iCCA cases (n = 7), 2% (0-3) in HCC cases (n = 5), and 64% (1-100) in cHCC-CCA cases (n = 8). The mean percentage of HCC IHC contingent was 4% (0-13) in iCCA cases, 76% (52-93) in HCC cases, and 17% (1-41) in cHCC-CCA cases. In the external validation set, most cHCC-CCA cases (5/6, 83%) presented a majority contingent of iCCA.

### Clustering results

After feature extraction, the validation set was clustered with a Gaussian mixture model. Different numbers of clusters were tested and analysed. The most coherent results were obtained for *K* = 2 clusters. In this two-cluster model, Cluster 0 contained tiles with homogeneous tissue and large cells with abundant eosinophil cytoplasm, features characteristic of HCC (76% of the tiles). The second cluster (Cluster 1) contained tiles with glandular structures within a fibrous stroma (stained orange in HES), usually present in iCCA tumours (92% of the tiles) (Fig. 6). Tiles from cHCC-CCA cases were spread among the two clusters.

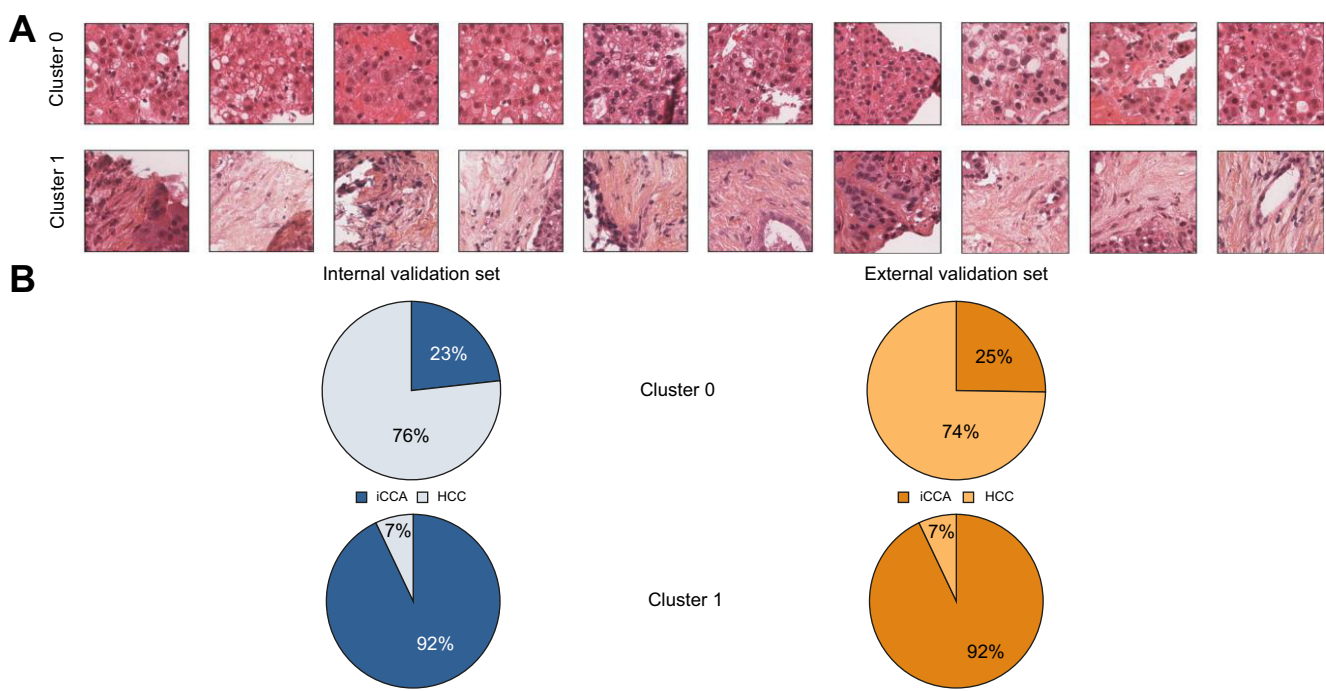

**Fig. 6. Results of the two-cluster model.** Examples of patches in Clusters 0 and 1 (A) and distribution of the HCC and iCCA tumour tiles within each cluster in the internal and external validation sets (B). HCC, hepatocellular carcinoma; iCCA, intrahepatic cholangiocarcinoma.

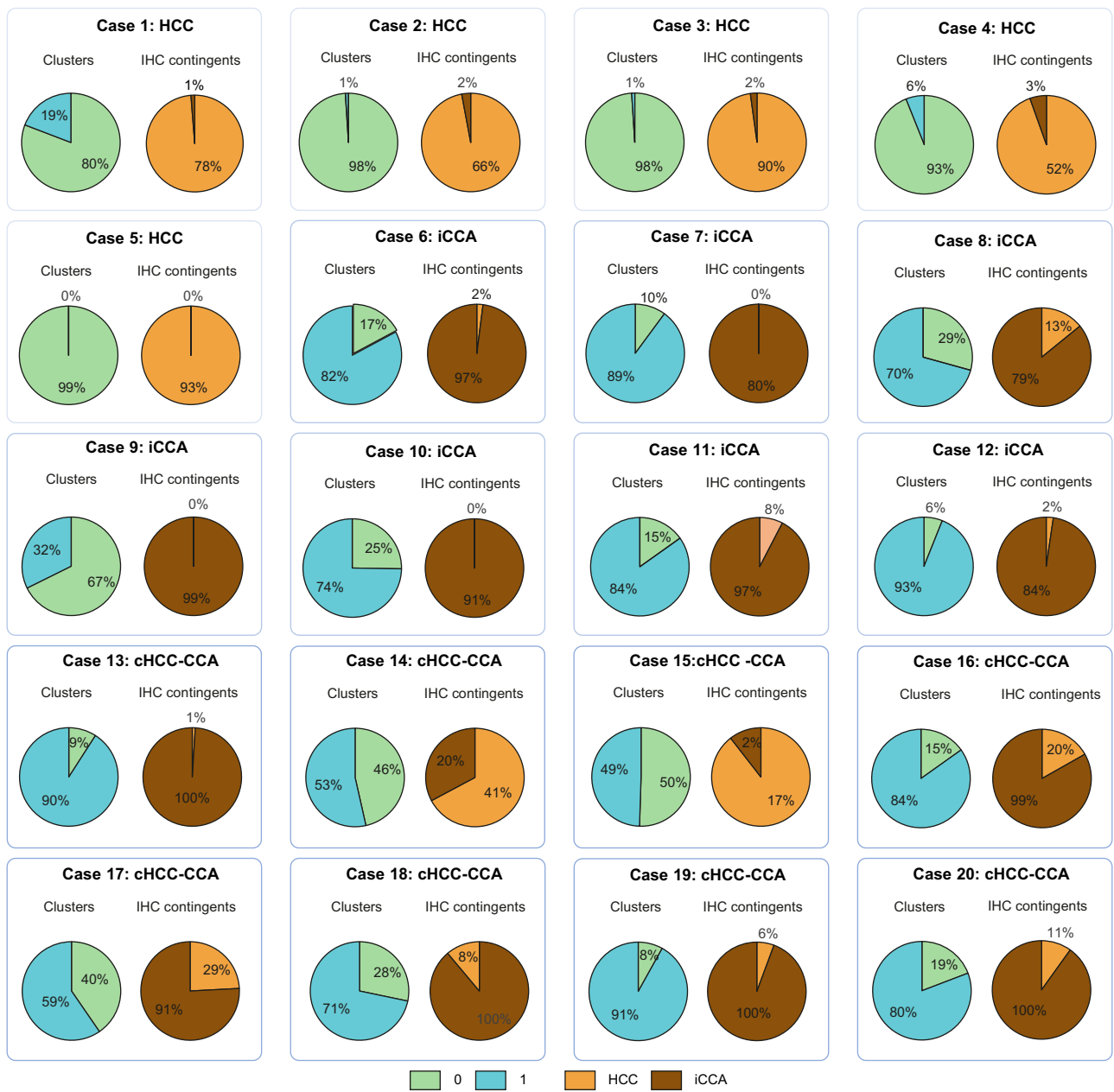

**Fig. 7. Comparison of the two-cluster predictions and immunohistochemistry contingents in the internal validation cohort (n = 20).** Pie charts showing (i) the proportion of Cluster 0 (green) and 1 (blue) tiles within each slide and (ii) the proportion of HCC (orange) and iCCA IHC contingents (brown) within each slide. The associated pathological diagnosis is displayed above each chart. cHCC-CCA, combined hepatocellular-cholangiocarcinoma; HCC, hepatocellular carcinoma; iCCA, intrahepatic cholangiocarcinoma.

We created a three-cluster model ($K = 3$) to test the potential of extracting a separate cluster for the combined cHCC-CCA tumour. The three-cluster model divided the previous Cluster 0 into two clusters (Clusters 2 and 3), with the previous Cluster 1 remaining separate (Cluster 4). The cHCC-CCA patches were spread among the three clusters. Half the cHCC-CCA cases (51%) were assigned to the cluster with a majority of iCCA (Cluster 4, corresponding to Cluster 1 in the two-cluster model), whereas the two new HCC clusters, Clusters 2 and 3, contained 26% and 31% of cHCC-CCA cases, respectively (Fig. S1).

### Internal validation of the two-cluster model
*Quantification of clusters in the validation set*
The two-cluster model was assessed with the validation set by quantifying the proportions of Cluster 0 (representative of HCC) and Cluster 1 (representative of iCCA) in each case. Overall, in validation set samples, the mean proportion of Cluster 0 was 51% (range 2-100%) and Cluster 1 was 48% (0-97%).

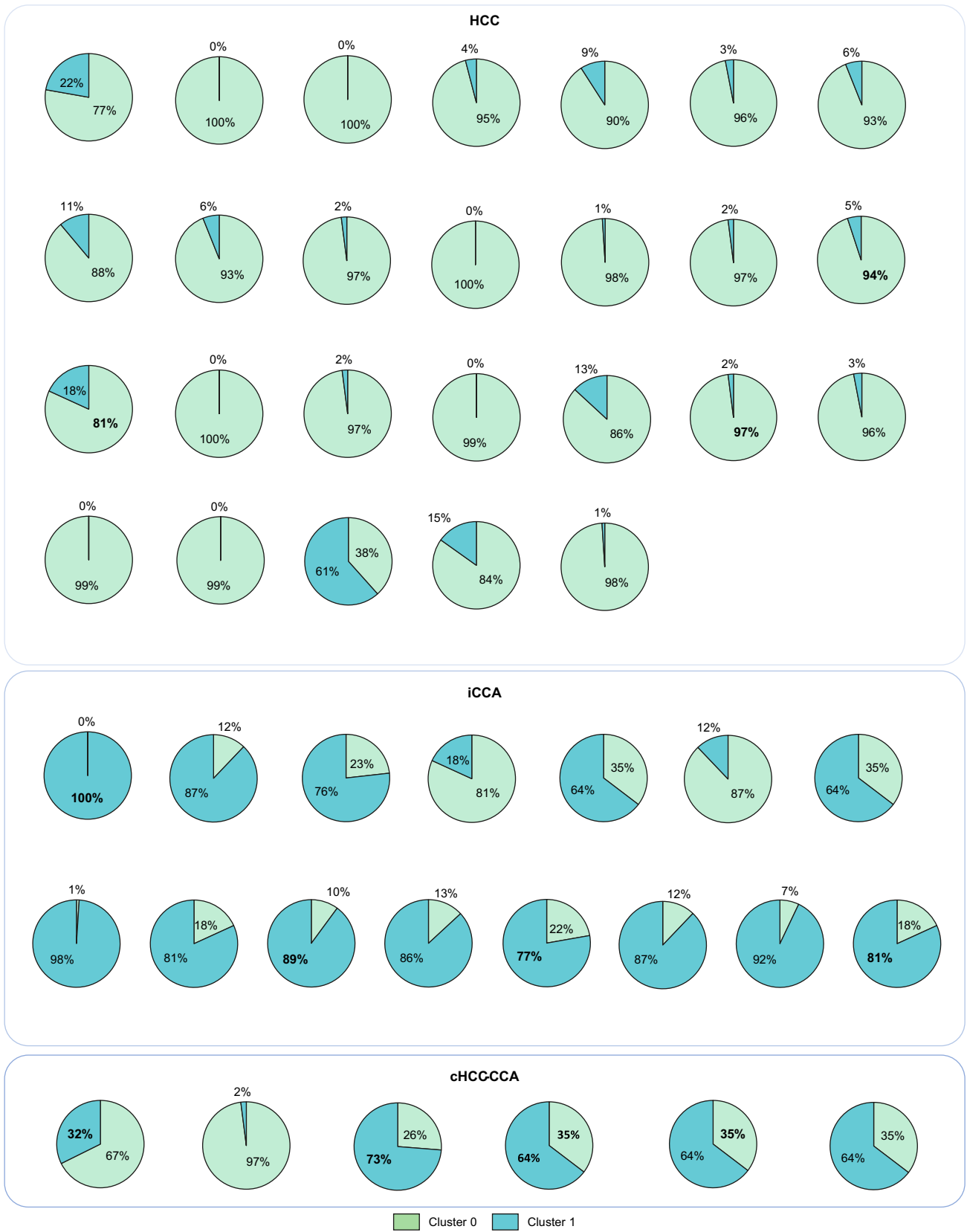

Fig. 8. Comparison of the two-cluster predictions and the pathological diagnosis within each slide of the external validation set (n = 47). Pie charts showing the proportion of Cluster 0 (green) and 1 (blue) tiles within each slide. The associated pathological diagnosis is displayed above each group. cHCC-CCA, combined hepatocellular-cholangiocarcinoma; HCC, hepatocellular carcinoma; iCCA, intrahepatic cholangiocarcinoma.

*Two-cluster model and pathology diagnosis*
For HCC and iCCA, the diagnostic agreement between the pathological diagnosis and the model predictions (major contingent) was 100% for HCC (11/11 cases) and 78% for iCCA (7/9 cases). For cHCC-CCA, we observed a highly variable proportion of each cluster type (Cluster 0 [5-50%] and Cluster 1 [9-94%]). The diagnostic agreement between the major contingent in conventional pathology and the one predicted by the model was 89% (8/9 cases) (Fig. S2).

*Two-cluster model and quantitative IHC analyses*
The diagnostic agreement between the IHC diagnosis and the two-cluster model predictions (major contingent) was 100% in HCC (5/5 cases), 86% in iCCA (6/7 cases) and 75% in cHCC-CCA (6/8 cases) (Fig. 7). The correlation coefficient between the proportions of Cluster 0 predicted by the two-cluster model and the HCC IHC contingent was 0.87, and between the proportions of Cluster 1 predicted by the two-cluster model and the iCCA IHC contingent was 0.77.

**External validation of the two-cluster model**
The results obtained with the two-cluster model for the internal validation set generalized well to the external validation set. Cluster 0 contained a majority of HCC tiles (74% of the tiles), and Cluster 1 contained mainly iCCA tiles (92% of the tiles) (Fig. 6). For HCC and iCCA, the diagnostic agreement between the pathological diagnosis and the model predictions (major contingent) was 96% (25/26) for HCC and 87% (13/15) for iCCA, respectively.

For cHCC-CCA, we observed a highly varying proportion of each cluster type (Cluster 0 [26-97%] and Cluster 1 [2-73%]). The diagnostic agreement between the major contingent in conventional pathology and the one predicted by the model was 83% (5/6 cases) (Fig. 8).

**Discussion**
This proof-of-concept study has shown that weakly supervised deep learning can extract discriminative features of different PLCs from routinely stained tumour biopsies. A two-cluster model based on Gaussian mixtures of these features has illustrated the discriminative power of biopsies and their potential for classifying different liver cancers.

At the core of the proposed approach was a combination of supervised transfer learning and unsupervised clustering. For the supervised part, we used a pre-trained convolutional neural network (ResNet18) with weak annotations of tumour/non-tumour areas in WSIs. As explained before, a complete annotation of these images would have required tremendous effort. Hence, a comparison with a fully supervised approach was impossible. Nonetheless, the adopted transfer learning approach successfully retrieved meaningful features with reasonable annotation effort and showed promising results for classifying PLC in biopsies.

Our model has identified two clusters, one specific to HCC and another specific to iCCA, corresponding to the two tumour types located at opposite ends of the malignant liver spectrum. The distinction between the two clusters aligns with the morphological variations between HCC, characterized by notable tumour eosinophilic hepatocytes without fibrous stroma, and iCCA, characterized by smaller cells with abundant fibrous stroma. Among HCC and iCCA cases in the validation set, two cases of iCCA were misclassified as HCC by the two-cluster model. A pathological review of these two cases revealed one moderately differentiated tumour and one poorly differentiated tumour, both exhibiting extensive architectural areas and minimal stroma. These factors could potentially explain the model's error. In the two-cluster model, cHCC-CCA tiles included varying proportions of both HCC and iCCA clusters. These results are consistent with the current WHO definition of the unequivocal presence of both hepatocytic and cholangiocytic differentiation within the same tumour on routine histopathology.[9]

The proposed two-cluster model provides the accurate percentage of HCC and iCCA components in cHCC-CCA, which could be valuable information for choosing a treatment strategy. Currently, there is no systemic treatment for unresectable cHCC-CCA, and patients with cHCC-CCA receive standard advanced iCCA or HCC treatments[28–31] without any specific recommendation. Hence, determining the main component of cHCC-CCA holds potential for guiding the best treatment strategy.

Our two-cluster model involved only routine HES slides. Although diagnosis is based on routine HES staining, additional IHC is still routinely performed to confirm the diagnosis of cHCC-CCA.[7,9] Comparing each contingent's percentage with the pathological diagnosis (based on both HES and IHC staining) showed that our HES only model achieved equivalent tumour classification. This consistency is remarkable as biopsy samples may be limited in size, preventing the use of complementary IHC. For example, 9 out of 29 cases (31%) in our cohort could not benefit from this additional analysis because of a shortage in tumour tissue. Interestingly, the model has identified HCC tiles within all iCCA cases, confirmed by IHC analysis in half of them, suggesting that cHCC-CCA could be more frequent than reported by the pathologist in the iCCA group. Thus, our model, providing a more accurate tumour tissue characterization than conventional histology, confirms previous findings showing tiny HCC or iCCA areas (not diagnosed by histology) within cHCC-CCA samples using MALDI imaging, an *in situ* proteomic approach.[32]

The potential therapeutic outcomes and the ability to circumvent material difficulties highlight the importance of developing automatic AI methodologies for biopsy specimens. Currently, most AI studies of PLC have focused on surgical samples,[13,14,19,33] but most patients do not have such samples during their entire cancer history, which introduces a selection bias. In contrast, tumour biopsies are increasingly performed in the context of PLC, which could be helpful for AI-based approaches. With this in mind, deep learning can be challenging with biopsies, as their representativeness may be limited for heterogeneous tumours, and their small sample size and intricate shape reduce the number of tiles useable for analysis. Artefacts such as fragmentation, folds and tears (less visible in surgical samples) can also hinder the use of biopsies. Nevertheless, a few studies have laid the groundwork for using biopsies for other cancers, particularly prostate cancer.[34–36] The present study confirms that biopsies can be exploited despite these challenges and indicates that encouraging deep learning-based results can be obtained for liver cancer. Our study promotes the use of biopsy for PLC diagnosis and supports recent results showing that tumour biopsy led to accurate diagnosis in 11% of nodules classified as LIRADS-5 on radiology.[37]

The main limitations of this work are its retrospective design and the relatively low number of studied cases. To the best of our knowledge, no open-access database of PLC biopsies is available. Moreover, HCC routine biopsies are scarce, and cHCC-CCA samples are even more so because of the extreme rarity of this subtype. In order to avoid unbalanced learning, we have selected a similar proportion of each tumour subtype in the training and internal validation sets, which has impacted the total number of studied cases. Additional prospective validation has yet to confirm the full potential of our model in assisting the pathologist.

A weakly supervised learning method is able to extract specific morphological features of HCC and iCCA from tumour biopsy. Despite no specific features of cHCC-CCA being recognized, the identification of HCC and iCCA tiles within a slide could facilitate the diagnosis of cHCC-CCA.

### Abbreviations
AI, artificial intelligence; cHCC-CCA, Combined hepato-cholangiocarcinoma; HES, hematein eosin saffron; HCC, hepatocellular carcinoma; IHC, Immunohistochemistry; iCCA, Intrahepatic cholangiocarcinoma; PLC, Primary liver cancer; WSI, Whole-slide image.

### Financial support
The authors received no financial support to produce this manuscript.

### Conflicts of interest
The authors declare no conflicts of interest that pertain to this work.

Please refer to the accompanying ICMJE disclosure forms for further details.

### Authors' contributions
Study concept and design (AB, KM, JG, JCP, VP), acquisition of data (AB, PEZ, NO, MA, GL); analysis and interpretation of data (AB, PEZ, NO, JCP, VP); drafting of the manuscript (AB, NO, KM); critical revision of the manuscript for important intellectual content (JCP, VP); statistical analysis (AB); study supervision (JCP, VP).

### Data availability statement
The datasets used and/or analysed during the current study are available from the corresponding author on reasonable request.

### Acknowledgement
We thank the Bernouilli Lab (Inria-APHP) for supporting this study.

### Supplementary data
Supplementary data to this article can be found online at https://doi.org/10.1016/j.jhepr.2024.101008.